\documentclass[preprint2]{aastex}
\usepackage{color}
\newcommand{\unit}[1]{\rm\,#1}
\newcommand{\iso}[2]{{}^{#1}{\rm #2}}

\title{
Effects of magnetic fields on the propagation of nuclear flames in magnetic white dwarfs
}
\author{
Masamichi Kutsuna \altaffilmark{1,2},
and 
Toshikazu Shigeyama \altaffilmark{1}
}
\altaffiltext{1}{
Research Center for the Early Universe,
Graduate School of Science, The University of Tokyo,
Bunkyo-ku, Tokyo 113-0033, Japan.
}
\altaffiltext{2}{
Department of Astronomy,
Graduate School of Science, The University of Tokyo,
Bunkyo-ku, Tokyo 113-0033, Japan.
}

\begin{document}


\shorttitle{Magnetic field effect on SNe Ia}
\shortauthors{KUTSUNA \& SHIGEYAMA}

\begin{abstract}
  We investigate effects of the magnetic field on the propagation of laminar flames of nuclear reactions taking place in white dwarfs (WDs) with the mass close to the Chandrasekhar limit.
  We calculate the velocities of laminar flames parallel and perpendicular to  uniform magnetic fields as eigenvalues of steady solutions for magnetic hydrodynamical equations.
  As a result, we find that  even when the magnetic pressure does not dominate the entire pressure it is possible for the magnetic field to suppress the flame propagation through the thermal conduction.
  Above the critical magnetic field, the flame velocity decreases with increasing magnetic field strength as $v \sim B^{-1}$.
  In media with densities of $10^{7}, 10^{8}, \,\mathrm{and}\,10^{9} \unit{g\,cm^{-3}}$,
  the critical magnetic fields are orders of $\sim 10^{10}, 10^{11}, \,\mathrm{and}\,10^{12} \unit{G}$,
  respectively.
\end{abstract}


\section{Introduction}
Type Ia supernovae (SNe Ia) are thought to be thermonuclear explosions of C+O white dwarfs (WDs) in close binary systems
\citep{1984ApJS...54..335I,1984ApJ...286..644N}.
If the mass accretion rate from the companion star is $\sim 10^{-7}\,M_\odot\unit{yr^{-1}}$, the WD can increase its mass to the Chandrasekhar limit ($1.38\,M_\odot$) and the evolution ends up with the ignition of carbon in the innermost region.
As a result, the explosion is expected to be regulated by this limit for all of this type of SNe to behave in a similar manner (ideally, in an identical manner).
Thus they are used as standard candles \citep{1998AJ....116.1009R, 1999ApJ...517..565P}.
In reality, observations of SNe Ia suggest that their peak luminosities have a significant diversity \citep[e.g.][]{1996AJ....112.2391H}.
Despite of this diversity, we can estimate the absolute magnitude of maximum light from the decline rate of the luminosity after the peak using a tight correlation between them \citep{1993ApJ...413L.105P}.
With this calibration method, SNe Ia are used as a cosmological distance indicator, which reveals the accelerating expansion of the Universe \citep{1998AJ....116.1009R, 1999ApJ...517..565P}.

Yet, we do not agree on the intrinsic mechanism for SNe Ia to generate their luminosity dispersion.
\citet{1996ApJ...457..500H} investigated different explosion scenarios including deflagrations and detonation models and compared their light curves with those of observed SNe Ia.
They found a correlation between the types of host galaxies and explosion mechanisms.
The metallicity in a progenitor star is suggested to affect the composition of the WD and influence the brightness of the resulting SN \citep{1998ApJ...495..617H, 1999ApJ...522L..43U, 2003ApJ...590L..83T}.
The double degenerate scenario, in which a binary system composed of double WD merges to a massive WD with the mass exceeding the Chandrasekhar limit to result in an explosion, naturally yields a variety of explosions depending on the total mass \citep{1984ApJS...54..335I}.
Another possibility is the effect of magnetic field in a progenitor WD  \citep{2004MNRAS.348..451G}.
The propagation of expanding flow generated by deflagration may be suppressed in the direction perpendicular to the magnetic field, which will cause an asymmetric explosion.
Thus a diversity of SNe Ia caused by different strengths of magnetic fields depends not only on the amount of synthesized $\iso{56}{Ni}$ but also on the lines of sight of observers.

Recent observations, especially Sloan Digital Sky Survey (SDSS) have discovered a number of WDs with strong surface magnetic fields greater than $10^4\unit{G}$ \citep{2003ApJ...595.1101S,2005AJ....130..734V}.
A significant fraction of them have polar magnetic field strengths ranging from $10^{6}\unit{G}$ to $10^{9}\unit{G}$. There is a tendency that more massive WDs have stronger magnetic fields.
Furthermore, WDs in close binary systems tend to have strong fields, which indicates that such magnetism is produced in common envelopes \citep{2008MNRAS.387..897T}.
Note that these field strengths are the surface values. At the center of a WD, it would be stronger.
Since the mass of observed magnetic WDs are significantly less than the Chandrasekhar limit. (the average mass is $\sim 0.9\,M_\odot$),
increasing their masses toward the Chandrasekhar limit, their fields are likely to become much stronger when these magnetic WDs explode as SNe Ia.

\citet{2004MNRAS.348..451G} studied effects of the magnetic field on the turbulent combustion front and calculated resulting asymmetry of the explosion by evaluating the growth rate of the Rayleigh-Taylor instability of the turbulent front suppressed by the existence of magnetic fields.

The effect of magnetic field is not only dynamical.
Electrons are trapped in the presence of the magnetic field, which suppresses the transverse thermal conduction.
The velocity of a laminar deflagration depends on the thermal conduction \citep{1959flme.book.....L}.
In the reacting region of a deflagration front, the diffusion timescale $\tau_{\rm diff} = l^2 / \kappa$, and the burning timescale $\tau_{\rm burn} = E / \dot{S}$ are comparable.
Here, $l$ is the mean free path of electrons, $\kappa$ the diffusion coefficient, $E$ the energy density, and $\dot{S}$ is the energy generation rate.
Equating these timescales, the flame velocity $v_\mathrm{fl}$ relative to the unburned medium is approximately estimated by the following formula,
\begin{equation}
  v_\mathrm{fl} \sim \frac{l}{\tau_{\rm diff}}
  \sim \sqrt{\kappa \frac{\dot{S}}{E}}
  \label{eq:v_approx}
\end{equation}
This suggests that the transverse laminar flame slows down in the presence of the magnetic field,  because the diffusion coefficient is proportional to the thermal conductivity.
Thus, this effect is needed to be taken into account for discussing possible mechanisms to give rise to an asymmetric explosion and the diversity of SNe Ia due to magnetic fields.

In this paper, we study the propagations of laminar flames in magnetic C+O WDs.
Since the width of the wave front is much smaller than the scale lengths of a WD, we consider flames propagating into uniform media with constant densities and uniform magnetic fields.
In this approximation,
we evaluate the critical magnetic field to start affecting the flame velocity for each density.

The structure of the paper is as follows. In Section 2, we present formulation for the steady laminar nuclear flame. Section 3 explains the numerical procedure to solve the equations satisfying  boundary conditions  and to obtain the propagation velocity as an eigenvalue problem.  Section 4 presents results of the calculations.
Section 5 concludes this paper.

\section{Formulation}

The laminar flame velocities in non-magnetic WDs are calculated by
\citet{1992ApJ...396..649T} (TW92).
They integrate time dependent one-dimensional hydrodynamical equations for laminar flames propagating in uniform media till they obtain approximately steady states. Then the velocity of the front is measured.

Instead, we integrate time independent one-dimensional steady state equations in the wave rest frame with a mass flux chosen so as to satisfy boundary conditions imposed at both ends of the flame front.
This method allows us to determine the velocity of the wave front as an eigenvalue of the steady solution.

\subsection{MHD equations}
Since we consider one-dimensional steady flows, physical variables including magnetic fields depend only on the spatial coordinate $x$.
MHD equations governing the flow perpendicular to the magnetic field can be written in the following manner.
\begin{eqnarray}
  \frac{d}{dx}  \left(\rho v \right) &=& 0,
  \label{eq:mass-perp}
  \\
  \frac{d}{dx} \left(\rho v^2 + P + \frac{B^2}{8\pi}\right) &=& 0,
  \label{eq:momentum-perp}
  \\
  \rho v \frac{d}{dx} \left(\frac{v^2}{2} + h + \frac{B^2}{4\pi\rho}\right)
  &=& \rho\dot{S} + \frac{d}{dx}\left(\lambda_{\rm n}\frac{dT}{dx}\right),
  \label{eq:energy-perp}
\end{eqnarray}
where 
$T$ is the temperature,
$\rho$ the density,
$v$ the velocity,
$P$ the pressure,
$h$ the specific enthalpy,
$\dot{S}$ the net energy generation rate including the heating process by nuclear burnings and the neutrino cooling.
Furthermore, the thermal conductivity $\lambda_{\rm n}$ is dependent on the magnetic field.
The MHD induction equation is supplemented as
\begin{equation}
  \frac{d}{dx} \left(v B\right) = 0.
\end{equation}
From the mass conservation law (\ref{eq:mass-perp}), 
We obtain the mass flux $m_{\rm n}$ as an invariant.
\begin{equation}
  m_{\rm n} \equiv \rho v = {\rm const}.
  \label{eq:massflux-perp}
\end{equation}

On the other hand, equations governing the flow along the magnetic field are listed as follows. 
\begin{eqnarray}
  \frac{d}{dx} \left(\rho v^2 + P\right) &=& 0,
  \label{eq:momentum-parallel}
  \\
  \rho v \frac{d}{dx} \left(\frac{v^2}{2} + h\right)
  &=& \rho\dot{S} + \frac{d}{dx}\left(\lambda_{\rm p}\frac{dT}{dx}\right),
  \label{eq:energy-parallel}
\end{eqnarray}
where $\lambda_{\rm p}$ is the thermal conductivity parallel to the magnetic field. The contributions of magnetic field in equations (\ref{eq:momentum-parallel}) and (\ref{eq:energy-parallel}) vanish because of the uniform magnetic field throughout the flame.

The mass flux becomes an invariant again, but takes a different value.   
\begin{equation}
  m_{\rm p} \equiv \rho v = {\rm const}.
  \label{eq:massflux-parallel}
\end{equation}

Because $\lambda_{\rm p}$ is not dependent on the magnetic field (see Section 2.3),
all the above equations are independent of the magnetic field.

We ignore the diffusion of ions. Therefore the Lewis number for each nuclei is assumed to be zero in the flames concerned here.

\subsection{Equation of state}

The equation of state used for this problem consists of ideal gas of ions, partially degenerate electrons together with electron positron pairs, and photons, which are in thermal equilibrium.
We make a numerical table of all thermal quantities as functions of chemical abundance of an isotope $i$ (or a nuclide $i$) $Y_i$, 
the degree of electron degeneracy $\eta$, and the temperature $T$.

\subsection{Thermal conduction}

We follow the procedure of  \citet{1999A&A...351..787P} for evaluations of thermal conductivities $\lambda_{\rm p}$ and $\lambda_{\rm n}$. 
Suppose the magnetic field lies along the $z$-axis, the electron thermal conductivities introduced in the previous sections can be expressed as components of a 2nd-rank tensor such as  
$\lambda_{\rm p} = (\lambda_{\rm e})_{zz}$ and $\lambda_{\rm n} = (\lambda_{\rm e})_{xx}$, because the collision time of electrons, which is dependent on the direction, becomes a tensor $\tau_{ij}$ given by
\begin{eqnarray}
  \tau_{zz} &=& \tau_0
  , \\
  \tau_{xx} &=& \tau_{yy} = \frac{\tau_0}{1 + (\omega\tau_0)^2}
  , \\
  \tau_{yx} &=& - \tau_{xy} = \frac{\omega\tau_0^2}{1 + (\omega\tau_0)^2}
  ,
\end{eqnarray}
where $\omega$ is the cyclotron frequency of an electron,
$\tau_0$ is the non-magnetic electron collision time. The other components are zero.
The tensorial conductivity $\lambda_{\rm e}$  in the non-quantizing magnetic field is written in terms of the following 2nd-rank tensors.
\begin{equation}
  \lambda_{\rm e} = k^2T (I_2 - I_1 I_0^{-1} I_1),
\end{equation}
where $k$ is the Boltzmann constant, and the tensors $I_n$ are expressed as
\begin{equation}
  (I_n)_{ij} = \int^\infty_0 \left(\frac{\mu - \epsilon}{k T}\right)^n
  \frac{p^3\tau_{ij}(\epsilon)}{3\pi^2 \hbar^3m^*_\mathrm{e}(\epsilon)}
  \left(-\frac{\partial f(\epsilon)}{\partial\epsilon}\right)
  \,d\epsilon,
\end{equation}
where
$\mu$ is the chemical potential of an electron,
$\epsilon$ the kinetic energy of an electron,
$\hbar$ the Planck constant,
$m_\mathrm{e}$ the rest mass of an electron,
and $p$ is the momentum.
The relation between $\epsilon$ and $p$ is
$(\epsilon+m_\mathrm{e}c^2)^2 = (m_\mathrm{e}c^2)^2 + (cp)^2$,
where $c$ is the speed of light.
The effective mass $m^*_\mathrm{e}$ of an electron has been introduced as $m^*_\mathrm{e}=m_\mathrm{e}+\epsilon/c^2$. $f(\epsilon)$ is the Fermi-Dirac distribution function,
\begin{equation}
  f(\epsilon) = 
  \frac{1}{1 + \exp\,(\epsilon - \mu)/{k T}}.
\end{equation}

In the limit of extremely degenerate electrons  \citep{1980SvA....24..425U},
$\lambda_{\rm e}$ is approximated by
\begin{equation}
  (\lambda_{\rm e})_{ij} \simeq \frac{\pi^2 n_e k^2 T}{3m^*_\mathrm{e}(\epsilon_{\rm F})}\,\tau_{ij}(\epsilon_{\rm F}),
\end{equation}
where
$n_e$ is the electron number density
and $\epsilon_{\rm F}$ is the electron Fermi energy.
The ratio of the thermal conductivities is expressed as 
\begin{equation}
  \frac{\lambda_{\rm n}}{\lambda_{\rm p}}
  \simeq
  \frac{1}{1 + (\omega\tau_0)^2}.
  \label{eq:thcond_ratio}
\end{equation}
Therefore the existence of a magnetic field reduces $\lambda_{\rm n}$ and $\lambda_{\rm n} < \lambda_{\rm p}$. As a consequence, Equation (\ref{eq:v_approx}) suggests that a magnetic field suppresses the transverse laminar flame velocity.

\subsection{Nuclear reaction}
We calculate a nuclear reaction network composed of 
the triple-$\alpha$ reaction,
($\alpha$,$\gamma$) reactions,
($\alpha$,p)(p,$\gamma$) reactions,
$\iso{12}{C}+\iso{12}{C}$, $\iso{12}{C}+\iso{16}{O}$, and $\iso{16}{O}+\iso{16}{O}$,
with relevant 25 isotopes from $\iso{1}{H}$ to $\iso{56}{Ni}$ (see Table~\ref{tab:isotope}).
This reaction network is smaller than the one used by TW92.
The reaction rates and the energy generation rates $\dot{S}$ are referred to REACLIB \citep[e.g.][]{2010ApJS..189..240C}.
Rate equations except for $\iso{4}{He}$ and $\iso{12}{C}$ are represented in the following form,
\begin{eqnarray}
  \frac{DY_i}{Dt}
  &=& v \frac{dY_i}{dx}
  \nonumber\\
  &=& \sum_{j}R_1{}^i_j(\rho, T)Y_j -  R_1{}^j_i(\rho, T)Y_i
  \nonumber\\
  &&+ \sum_{j,k} \left[\frac{1}{1+\delta_{jk}}R_2{}^i_{jk}(\rho, T)Y_jY_k \right.\nonumber\\
  &&\left. - R_2{}^k_{ij}(\rho, T)Y_iY_j \right]
  \label{eq:reaction}
  ,
\end{eqnarray}
where $Y_i$ is the abundance of a nuclide $i$ with the mass number $A_i$ related to the mass fraction $X_i$ by $Y_i=X_i/A_i$, , $R_1{}_{i}^{j}$ the decay rate of a nuclide $i$ to a nuclide $j$, and $R_2{}^k_{ij}$ is the rate of the reaction between nuclide $i$ and $j$ to produce a nuclide $k$.
The rate equations for $\iso{4}{He}$ and $\iso{12}{C}$ include contributions from the triple-$\alpha$ reaction in addition to the right hand side of equation (\ref{eq:reaction}).
Additionally, we considered the electron screening effects by \citet{1990ApJ...362..620I}.

\section{Boundary conditions}

We solve the MHD equations in each direction
(eq. [\ref{eq:mass-perp}]-[\ref{eq:energy-perp}] for the transverse direction
and eq. [\ref{eq:momentum-parallel}]-[\ref{eq:massflux-parallel}] for the longitudinal direction), 
with the rate equations (\ref{eq:reaction}) satisfying the following boundary conditions.

As boundary conditions in the upstream, we assume the state of an unburned WD material.
The initial composition of the WD is assumed to be $X(\iso{12}{C})=0.5,\ X(\iso{16}{O})=0.5$.
For a given density, we set the initial temperature to the value at which the heating rate due to the nuclear burning is equal to the neutrino cooling rate \citep{1996ApJS..102..411I}.
The other boundary conditions $d/dx = 0$ is required at $x \rightarrow \infty$.
This is an eigenvalue problem for the mass flux $m$ that satisfies the boundary conditions at the two points.
We seek the value of $m$ to satisfy all the boundary conditions by integrating the MHD equations with various values of $v_0$, the velocity at $x=0$ in the wave rest frame.
When a trial value of $m$ is too large, the temperature increases faster with increasing $x$ than the true solution, eventually diverges, and never reaches $x \rightarrow \infty$.
On the other hand, when $m$ is too small, the temperature eventually decreases to zero before reaching $x\rightarrow\infty$.
Thus we can find the true value of $m$ between values of these two trends, and the corresponding $v_0$ becomes the flame velocity.

We obtain solutions with the densities in the unburned media $\rho_0$ ranging from $10^7$ to $10^9 \unit{g\,cm^{-3}}$.

\section{Results}
\subsection{Flame propagation along the magnetic field}
First, we present results for flames propagating along the magnetic field, which is equivalent to flames in non-magnetic media. We make a comparison of the resultant flame velocities with those derived by TW92. 

\begin{figure}
  \plotone{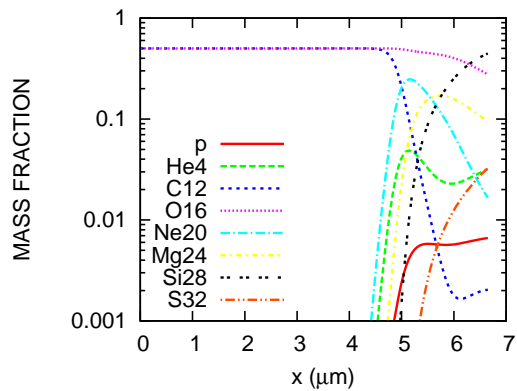}

  \caption{
  Distributions of chemical abundance of a flame parallel to the magnetic field, with
  $\rho_0 = 10^9\unit{g\,cm^{-3}}$.
  \label{fig:parallel.X}
  }
\end{figure}

When the initial density is $\rho_0 = 10^9\unit{g\,cm^{-3}}$, the eigen value is found to be $m_\mathrm{p} \simeq 3.1 \times 10^{15}\unit{g\,cm^{-2}\,s^{-1}}$, and the flame velocity is $v_0 \simeq 31 \unit{km\,s^{-1}}$.
Figure~\ref{fig:parallel.VT} shows the resultant profiles of density and temperature.
The result of nuclear abundances is presented in Figure~\ref{fig:parallel.X}. The carbon burning leads to the production of elements up to sulfur, which is consistent with the result of TW92.

\begin{figure}
  \plotone{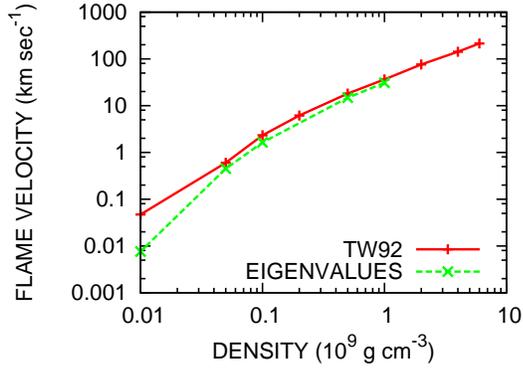}

  \caption{
  Laminar flame velocities for different densities in the unburned region.
  \label{fig:timmes}
  }
\end{figure}

In Figure~\ref{fig:timmes}, the flame velocities are compared with the result of TW92.
The flame velocities of our results  are smaller than those of TW92.
In TW92, a smaller reaction network (including fewer number of nuclei and reactions) resulted in smaller flame velocities. Since TW92 used a reaction network larger than our network, the smaller flame velocities of our results may be partly due to this difference.
It is also claimed in TW92 that the velocity derived from the eigenvalue method is smaller than from the dynamical equations.
The difference is prominent at low densities. 
Different reaction rates and thermal conductivities in our calculations might be another factor to deviate the flame speeds from those of TW92.

\subsection{Flame propagation across the magnetic field}
\begin{figure}
  \plotone{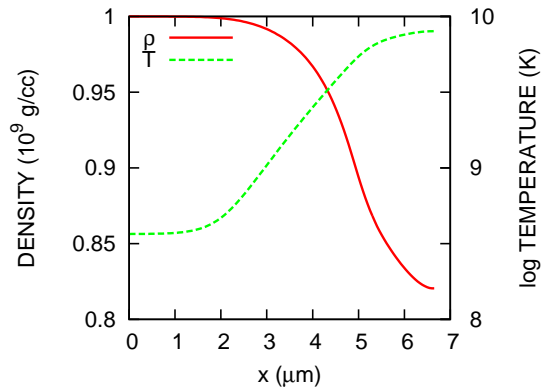}

  \caption{
  Temperature and density profile of a flame parallel to the magnetic field, with
  $\rho_0 = 10^9\unit{g\,cm^{-3}}$.
  The eigenvalue $m_\mathrm{p} \simeq 3.1 \times 10^{15}\unit{g\,cm^{-2}\,s^{-1}}$, and the flame velocity $v_0 \simeq 31 \unit{km\,s^{-1}}$.
  \label{fig:parallel.VT}
  }
\end{figure}

\begin{figure}
  \plotone{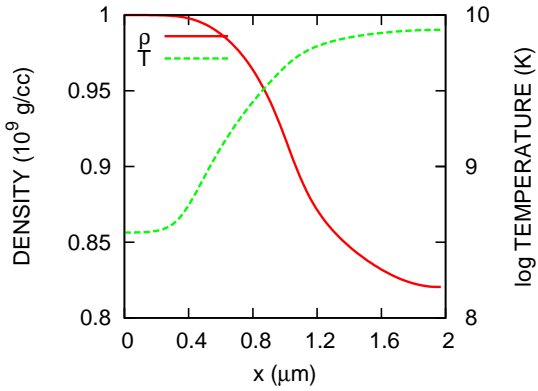}

  \caption{
  Temperature and density profile of a flame perpendicular to the magnetic field, with
  $\rho_0 = 10^9\unit{g\,cm^{-3}}$,
  $B_0 = 10^{12}\unit{G}$.
  The eigenvalue $m_\mathrm{n} \simeq 1.8 \times 10^{15}\unit{g\,cm^{-2}\,s^{-1}}$, and the flame velocity $v_0 \simeq 18 \unit{km\,s^{-1}}$.
  \label{fig:perp.VT}
  }
\end{figure}

\begin{figure}
  \plotone{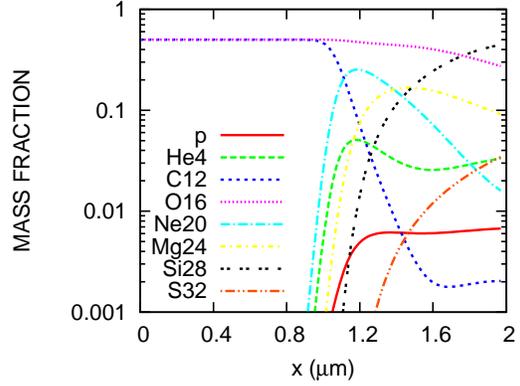}

  \caption{
  Distributions of chemical abundance of a flame perpendicular to the magnetic field, with
  $\rho_0 = 10^9\unit{g\,cm^{-3}}$,
  $B_0 = 10^{12}\unit{G}$.
  \label{fig:perp.X}
  }
\end{figure}

Figures~\ref{fig:perp.VT} and \ref{fig:perp.X} show the solution for a flame propagating in the direction perpendicular to the magnetic field with the initial density $\rho_0 = 10^9\unit{g\,cm^{-3}}$ and the initial magnetic field $B_0 = 10^{12}\unit{G}$.
The corresponding eigen value becomes $m_\mathrm{n} \simeq 1.8 \times 10^{15}\unit{g\,cm^{-2}\,s^{-1}}$, and the flame velocity $v_0 \simeq 18 \unit{km\,s^{-1}}$.
The scale of the burning region becomes shorter than that of the flame presented in Figures~\ref{fig:parallel.VT} and \ref{fig:parallel.X}.
The velocity becomes slower due to the reduction of the thermal conductivity by the magnetic field.
Both of the scale and the velocity decrease with increasing $B_0$.

\begin{figure}
  \plotone{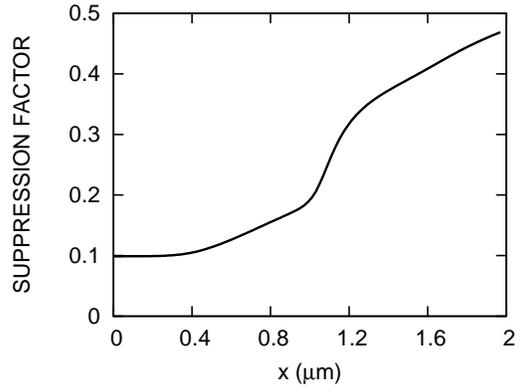}
  \caption{
  Ratio of conductivities $\lambda_\mathrm{n}/\lambda_\mathrm{p}$ as a function of position $x$ for a flame propagating normal to the magnetic field.
  Here $\lambda_\mathrm{n}$ is the conductivity in the direction normal to the magnetic field and $\lambda_\mathrm{p}$ is evaluated by using the thermal state at each position and neglecting the magnetic effects.
  $\rho_0 = 10^9\unit{g\,cm^{-3}}$,
  $B_0 = 10^{12}\unit{G}$.
  \label{fig:thcond.ratio}
  }
\end{figure}

To see where the magnetic field affects the propagation of a flame normal to the magnetic field, Figure~\ref{fig:thcond.ratio} shows the distribution of the ratio of conductivities $\lambda_\mathrm{n}/\lambda_\mathrm{p}$.
Here $\lambda_\mathrm{n}$ is the conductivity in the direction normal to the magnetic field and $\lambda_\mathrm{p}$ is evaluated by using the thermal state at each position and neglecting the magnetic effects.
The velocity of the flame is suppressed by a factor of $\sim 2$ due to the magnetic field (see Figs. (\ref{fig:parallel.VT}) and (\ref{fig:perp.VT})).
This means the suppression of the conductivity by a factor of $\sim4$ from Equation (\ref{eq:v_approx}).
Therefore Figure~\ref{fig:thcond.ratio}  indicates that the magnetic field in the region where carbon burning takes place determines the flame velocity.

\subsection{Critical magnetic fields}
\begin{figure}
  \plotone{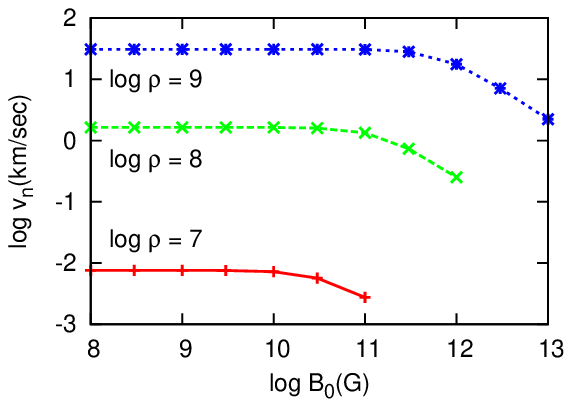}
  \caption{
  Velocities of laminar flames propagating across the magnetic fields
  as functions of field strengths for densities of
  $\rho_0 = 10^{7},\,10^8,\,10^9\unit{g\,cm^{-3}}$.
  \label{fig:B-V}
  }
\end{figure}

We calculate the flame velocities $v_{\rm n}$ normal to the magnetic fields for various strengths $B_0$ of fields and densities $\rho_0$. 
Figure~\ref{fig:B-V} shows the dependence of the flame velocities $v_{\rm n}$ on the strength of magnetic field for a few different densities.
Equations (\ref{eq:v_approx}) and (\ref{eq:thcond_ratio}) imply $v_{\rm n} / v_{\rm p} \sim B_0^{-1}$ in the limit of large $B_0$.
Figure~\ref{fig:B-V} shows this trend for each density.
If we define the critical field strength $B_\mathrm{cr}$ as the strength of the magnetic field that suppresses the flame velocity $v_{\rm n}$ by a factor of 2,
the critical magnetic fields are estimated as $B_{\rm cr} \sim 10^{10}, 10^{11}, \,\mathrm{and}\,10^{12} \unit{G}$ for densities $\rho_0 = 10^{7}, 10^{8}, \,\mathrm{and}\,10^{9} \unit{g\,cm^{-3}}$, respectively.
Therefore  even when the magnetic pressure does not dominate entire pressure, it is possible that the magnetic field suppresses the flame propagation through the thermal conduction.

\begin{figure}
  \plotone{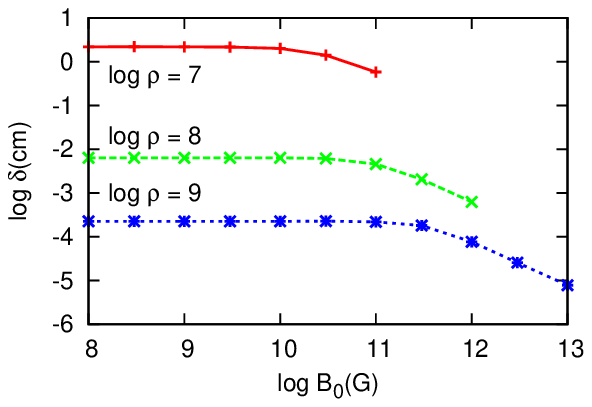}
  \caption{
  Widths of laminar flames propagating across the magnetic fields
  as functions of field strengths for densities of
  $\rho_0 = 10^{7},\,10^8,\,10^9\unit{g\,cm^{-3}}$.
  \label{fig:B-W}
  }
\end{figure}

Figure~\ref{fig:B-W} shows the dependence of the flame width $\delta = (T_{\rm b} - T_{\rm u}) / |dT/dx|_{\rm max}$ on the strength of magnetic field.
Here, $T_{\rm b}$ and $T_{\rm u}$ are temperatures in burned and unburned media.  $|dT/dx|_{\rm max}$ is the maximum value of temperature gradients in the wave front.
The flame width decreases when the strength of magnetic fields increases above the critical value, which is the same trend as $v_{\rm n}$ shows.
This implies that the burning timescale ($\sim \delta/v_{\rm n}$) is not affected by the existence of a magnetic field.

\section{Conclusions and discussion}
We study the propagation of the laminar flames in MWDs.
We calculate the laminar flame velocities parallel and perpendicular to the magnetic field as the eigensolutions of steady MHD equations taking into account nuclear reactions and thermal conduction by electrons.
From the results, we estimate the critical magnetic fields $B_{\rm cr}$ for given densities in the upstream region relevant to the interior of a C+O WD. Consequently, we find that magnetic fields of the order of $B_{\rm cr}\sim 10^{12}\unit{G}$ significantly suppress the flame velocity at the center of the WD (the density thereof is $\rho \sim 10^9\unit{g\,cm^{-3}}$) with the mass close to the Chandrasekhar limit.

Though only the electron thermal conduction is concerned in our calculations, photons also transport the heat generated by nuclear reactions.
The photon thermal conductivity increases with decreasing densities and increasing temperatures.
On the other hand, the electron conductivity is not sensitive to these quantities, because of the high degeneracy.
\citet{2007ApJ...655L..93C} calculated the propagation of nuclear flames in non-magnetic media with the thermal conduction including photons as well as electrons.
They found that photons are more efficient than electrons for the heat transport in lower density media ($\rho < 7 \times 10^8\unit{g\,cm^{-3}}$).

Recent observations have discovered many MWDs with strong surface magnetic fields up to $1\times10^9\unit{G}$.
The mass of the MWD with the strongest magnetic field discovered so far is $\sim 0.9\,M_\odot$.
The corresponding central density is estimated to be $\rho_c \sim 10^7\unit{g\,cm^{-3}}$.
If this MWD were to increase its mass close to the Chandrasekhar limit, carbon would be ignited at the center with the density of $\sim 10^9\unit{g\,cm^{-3}}$.
Meanwhile the strength of the magnetic field would increase scaling with the density as  $B\sim\rho^{2/3}$ unless the electrical resistance dissipated the magnetic energy. Thus $B$ will be enhanced by a factor of $\sim 20$.
According to \citet{2006ApJS..164..156Y}, there exist equilibrium configurations of non-rotating magnetized polytropic gas with the index $N=3$ that has the ratio of the field strengths at the center to the pole as high as $\sim$45.
Therefore the magnetic field is expected to become $20\times 45\times 10^9 \sim 10^{12}\unit{G}$ at the center, comparable to the critical strength.
In addition, a pre-supernova WD has experienced convection heated by $\iso{12}{C} + \iso{12}{C}$ reactions \citep{2000ARA&A..38..191H}.
It will amplify the local magnetic field, which makes it more likely that the magnetic field exceeds the critical strength.

\citet{2004MNRAS.348..451G} calculated the turbulent flame velocities propagating in the MWD.
They considered effects of magnetic fields only on the growth rate of the Rayleigh-Taylor instabilities. The turbulent flame velocity was estimated by fractal scaling.
Here the laminar flame velocity corresponds to the maximum wave length  of the Rayleigh-Taylor instability and the turbulent flame velocity the minimum wave length.
The minimum wave length depends on the magnetic field through the growth rate of the instability, while the laminar flame velocity was assumed to be independent of the magnetic field.
We find that the laminar flame velocity is also affected by magnetic fields even weaker than those that can change the growth rate of the Rayleigh-Taylor instability.
If this effect found in this paper is taken into account, the procedure of \citet{2004MNRAS.348..451G} would suggest more asymmetric explosions.
In this approximation, magnetic fields slow down the nuclear burning.
Therefore the MWD can expand to a larger extent before the flame reaches the surface. As a result, the nuclear burning occurs in a lower density environment, which leads to the production of a less amount of $\iso{56}{Ni}$.
However, there are some non-linear effects that may lead to significantly different results.
For example, there must be a feed-back of a suppressed flame velocity perpendicular to the magnetic field to the thermal conduction along the field.
The residual heat from the nuclear burning that cannot be conducted in the direction perpendicular to the field may be conducted along the field and change the flame velocity in this direction.
Change of nuclear burning may alter the expansion of the WD compared to a non-magnetic explosion.
To investigate such effects, we need to perform multi-dimensional calculations, which is beyond the scope of the present paper.

\begin{table}
  \caption{Nuclear isotopes contained in our reaction network. The top pannel shows reactants and products of $(\alpha,\gamma)$ reactions. The bottom pannel shows those of $(\alpha,{\rm p})$ reactions. \label{tab:isotope}}
  \begin{tabular}{|c|p{0.7\linewidth}|}
    \tableline
    $(\alpha,\gamma)$ &
    $\alpha$, $\iso{12}{C}$, $\iso{16}{O}$, $\iso{20}{Ne}$, $\iso{24}{Mg}$, $\iso{28}{Si}$, $\iso{32}{S}$, $\iso{36}{Ar}$, $\iso{40}{Ca}$, $\iso{44}{Ti}$, $\iso{48}{Cr}$, $\iso{52}{Fe}$, $\iso{56}{Ni}$
    \\
    \tableline
    $(\alpha,{\rm p})$ & 
    ${\rm p}$, $\iso{15}{N}$, $\iso{19}{F}$, $\iso{23}{Na}$, $\iso{27}{Al}$, $\iso{31}{P}$, $\iso{35}{Cl}$, $\iso{39}{K}$, $\iso{43}{Sc}$, $\iso{47}{V}$, $\iso{51}{Mn}$, $\iso{55}{Co}$ \\
    \tableline
\end{tabular}
\end{table}

\begin{thebibliography}{}
  \bibitem[Chamulak et al.(2007)]{2007ApJ...655L..93C} Chamulak, D.~A., 
    Brown, E.~F., \& Timmes, F.~X.\ 2007, \apjl, 655, L93
  \bibitem[Cyburt et al.(2010)]{2010ApJS..189..240C} Cyburt, R.~H., et al.\ 
    2010, \apjs, 189, 240 
  \bibitem[Ghezzi et al.(2004)]{2004MNRAS.348..451G}
    Ghezzi, C.~R., de Gouveia Dal Pino, E.~M., \& Horvath, J.~E.\ 2004, \mnras, 348, 451
\bibitem[Hamuy et al.(1996)]{1996AJ....112.2391H} Hamuy, M., Phillips, 
M.~M., Suntzeff, N.~B., et al.\ 1996, \aj, 112, 2391 
    
  \bibitem[Hillebrandt 
    \& Niemeyer(2000)]{2000ARA&A..38..191H} Hillebrandt, W., \& Niemeyer, J.~C.\ 2000, \araa, 38, 191 
  \bibitem[Hoeflich 
    \& Khokhlov(1996)]{1996ApJ...457..500H} Hoeflich, P., \& Khokhlov, A.\ 1996, \apj, 457, 500 
  \bibitem[Hoeflich et al.(1998)]{1998ApJ...495..617H} Hoeflich, P., Wheeler, 
    J.~C., \& Thielemann, F.~K.\ 1998, \apj, 495, 617 
  \bibitem[Iben 
    \& Tutukov(1984)]{1984ApJS...54..335I} Iben, I., Jr., \& Tutukov, A.~V.\ 1984, \apjs, 54, 335 
  \bibitem[Itoh et al.(1996)]{1996ApJS..102..411I} Itoh, N., Hayashi, H., 
    Nishikawa, A., \& Kohyama, Y.\ 1996, \apjs, 102, 411
  \bibitem[Itoh et al.(1990)]{1990ApJ...362..620I} Itoh, N., Kuwashima, F., 
    \& Munakata, H.\ 1990, \apj, 362, 620
  \bibitem[Landau 
    \& Lifshitz(1959)]{1959flme.book.....L} Landau, L.~D., \& Lifshitz, E.~M.\ 1959, Course of theoretical physics, Oxford: Pergamon Press, 1959,  
  \bibitem[Nomoto et al.(1984)]{1984ApJ...286..644N} Nomoto, K., Thielemann, 
    F.-K., \& Yokoi, K.\ 1984, \apj, 286, 644 
  \bibitem[Perlmutter et al.(1999)]{1999ApJ...517..565P} Perlmutter, S., et 
    al.\ 1999, \apj, 517, 565 
  \bibitem[Phillips(1993)]{1993ApJ...413L.105P} Phillips, M.~M.\ 1993, \apjl, 
    413, L105 
  \bibitem[Potekhin(1999)]{1999A&A...351..787P} Potekhin, A.~Y.\ 1999, \aap, 351, 787 
  \bibitem[Riess et al.(1998)]{1998AJ....116.1009R} Riess, A.~G., et al.\ 
    1998, \aj, 116, 1009
  \bibitem[Schmidt et al.(2003)]{2003ApJ...595.1101S} Schmidt, G.~D., et al.\ 
    2003, \apj, 595, 1101 
  \bibitem[Timmes \& Woosley(1992)]{1992ApJ...396..649T} Timmes, F.~X., \& Woosley, S.~E.\ 1992, \apj, 396, 649 
  \bibitem[Timmes et al.(2003)]{2003ApJ...590L..83T} Timmes, F.~X., Brown, 
    E.~F., \& Truran, J.~W.\ 2003, \apjl, 590, L83 
  \bibitem[Tout et al.(2008)]{2008MNRAS.387..897T} Tout, C.~A., 
    Wickramasinghe, D.~T., Liebert, J., Ferrario, L., 
    \& Pringle, J.~E.\ 2008, \mnras, 387, 897 
  \bibitem[Umeda et al.(1999)]{1999ApJ...522L..43U} Umeda, H., Nomoto, K., 
    Kobayashi, C., Hachisu, I., \& Kato, M.\ 1999, \apjl, 522, L43 
  \bibitem[Urpin \& Yakovlev(1980)]{1980SvA....24..425U} Urpin, V.~A., \& Yakovlev, D.~G.\ 1980, \sovast, 24, 425 
  \bibitem[Vanlandingham et al.(2005)]{2005AJ....130..734V} Vanlandingham, 
    K.~M., et al.\ 2005, \aj, 130, 734
  \bibitem[Yoshida 
    \& Eriguchi(2006)]{2006ApJS..164..156Y} Yoshida, S., \& Eriguchi, Y.\ 2006, \apjs, 164, 156 

\end{thebibliography}
\end{document}